\documentclass[twocolumn]{aastex63}

\newcommand{\arf}[1]{\texttt{arf}}
\newcommand{\rmf}[1]{\texttt{rmf}}

\received{June 1, 2019}
\revised{January 10, 2019}
\accepted{\today}
\submitjournal{AJ}

\shorttitle{}
\shortauthors{Rhea et al.}
\usepackage{amsmath}

\DeclareMathOperator*{\argmin}{arg\,min}
\graphicspath{{./}{figures/}}
\usepackage{comment}
\usepackage{physics}
\begin{document}

\title{A Data-driven Approach to X-ray Spectral Fitting: Quasi-Deconvolution}

\correspondingauthor{Carter Rhea}
\email{carterrhea@astro.umontreal.ca}

\author[0000-0003-2001-1076]{Carter Rhea}
\affiliation{Département de Physique, Université de Montréal, Succ. Centre-Ville, Montréal, Québec, H3C 3J7, Canada}
\affiliation{Centre de recherche en astrophysique du Québec (CRAQ)}

\author[0000-0001-7271-7340]{Julie Hlavacek-Larrondo}
\affiliation{Département de Physique, Université de Montréal, Succ. Centre-Ville, Montréal, Québec, H3C 3J7, Canada}

\author[0000-0002-0765-0511]{Ralph Kraft}
\affiliation{Smithsonian Astrophysical Observatory, Cambridge, MA 02138, USA}

\author{Akos Bogdan}
\affiliation{Smithsonian Astrophysical Observatory, Cambridge, MA 02138, USA}

\author{Rudy Geelen}
\affiliation{Oden Institute for Computational Engineering and Sciences, University of Texas at Austin, Austin, TX 78712, USA}

\begin{abstract}
X-ray spectral fitting of astronomical sources requires convolving the intrinsic spectrum or model with the instrumental response. Standard forward modeling techniques have proven success in recovering the underlying physical parameters in moderate to high signal-to-noise regimes; however, they struggle to achieve the same level of accuracy in low signal-to-noise regimes. Additionally, the use of machine learning techniques on X-ray spectra requires access to the intrinsic spectrum. Therefore, the measured spectrum must be effectively deconvolved from the instrumental response. In this note, we explore numerical methods for inverting the matrix equation describing X-ray spectral convolution. We demonstrate that traditional methods are insufficient to recover the intrinsic X-ray spectrum and argue that a novel approach is required.
\end{abstract}

\section{Introduction} \label{sec:intro}
The intrinsic emission spectrum of an X-ray source, $F(E)$, is defined by the underlying physical emission processes over a continuum of photon energies, $E$ (e.g. \citealt{kahn_direct_1980}; \citealt{weisskopf_chandra_1999}; \citealt{weisskopf_overview_2002}). The spectrum has units of  photons m$^{-2}$ s$^{-1}$ keV$^{-1}$. 
We relate it to the observed spectrum, $S(E')$, through a convolution with the instrumental response, $R(E',E)$:
\begin{equation}\label{eqn:spec}
    S(E') = \int_0^\infty R(E',E)F(E)dE
\end{equation}
Note that $E'$ denotes the discrete photon energies captured by the detector. Fitting over a broadband spectrum in this manner is common practice in X-ray astronomy due to the limited resolution of the detectors.
The instrumental response matrix for the \textit{Chandra X-ray Observatory} is captured in two parts: the redistribution matrix file (\rmf{}) and the ancillary response file (\arf{}) which are encapsulated in $R(E',E)$. The \rmf{} contains the mapping from the continuous energy space to the detector position space. Analogously the \arf{} contains the effective area of the detector as well as its quantum efficiency as a function of time-averaged energy. Although we focus on \textit{Chandra} response matrices, they are ubiquitous in X-ray astronomy; thus, this problem extends to other existing and future X-ray observatories.
The standard method for determining the true spectrum requires the following prescription: choose an appropriate parametric, physically-derived model to explain the intrinsic emission and fit the model using equation \ref{eqn:spec}. The fit is generally optimized by reducing the chi-squared statistic (e.g. \citealt{arnaud_xspec_1996}). 

Alternatively, it is possible to deconvolve equation \ref{eqn:spec} and directly solve for $F(E)$. However, the ill-conditioning of the response matrix makes this method unstable and thus infrequently used (e.g. \citealt{blissett_restoration_1979}). Matrices are considered ill-conditioned when their rows are not linearly independent of one another.
Certain applications, such as extracting spectral parameters using machine learning techniques (e.g. Rhea et al. 2021b), require the intrinsic spectrum rather than the observed spectrum. Therefore, the response matrix must be taken into account in order to isolate $F(E)$ from equation \ref{eqn:spec}.
Since the response matrix greatly affects the observed spectrum and changes significantly across the \textit{Chandra} field-of-view, the handling of it is crucial to proper analysis. 
Thus, we investigate several numerical methods to deconvolve the intrinsic spectrum from the response matrix.


\section{Methodology} \label{sec:meth}
In the following section, we will describe the derivation of (and reasoning behind) the matrix formulation of equation \ref{eqn:spec} and methods for solving the matrix equation.
\subsection{Deriving the Matrix Formulation}
Despite the simplicity of equation \ref{eqn:spec}, a direct convolution of the instrumental response function and model spectrum poses several issues. 
Due to the finite spectral resolution, the rows of the response matrix are not independent.
We confine the integral to the energies covered by the detector. Additionally, since the sampling of the detector space $E'$ is discrete, we can rewrite equation \ref{eqn:spec} as a matrix equation (\citealt{kaastra_optimal_2016}):
\begin{equation}\label{eqn:spec_mat}
    S_i = \sum_{j}R_{ij}F_j
\end{equation}
We have replaced the instrumental response function by its matrix counterpart, $R_{ij}$. 
In this formulation, $S_i$ represents the observed photon count rate in units of counts s$^{-1}$ for a given detector energy bin. $F_j$ is the model spectrum flux in units of counts m$^{-2}$ s$^{-1}$ in emitted energy bin $j$. For simplicity, we will write equation \ref{eqn:spec_mat} in the following form:
\begin{equation}\label{eqn:sim_mat}
    \vb{R}*\va{f} = \va{s}
\end{equation}

\subsection{Solution Methods}

Several methods for standard matrix equation solutions exist ($Ax=b$; ref); however, our application poses an additional constraints: the response matrix is ill-condition (condition number $>> 100$; e.g. \citealt{wilkinson_note_1972}).
The most straight-forward solution is to use a Moore-Penrose pseudo-inverse (e.g. \citealt{penrose_generalized_1955}) to invert the response matrix. In doing so we can directly calculate the intrinsic spectrum with a single matrix multiplication:
\begin{equation}\label{eqn:pseudo}
    \va{f} = \vb{R^{\dagger}}\va{s}
\end{equation}
where $\vb{R^\dagger}$ is the pseudo-inverse. The pseudo-inverse allows for the computation of an unique inverse matrix for non-square systems. Although this method is computationally efficient and direct, it suffers spurious oscillations owing to the ill-conditioning of $\vb{R}$ (e.g. \citealt{varah_numerical_1973}). In order to diminish the effects, we instead solve the normal equations:
\begin{equation}
    (\vb{R}^{T}\vb{R})\va{f} = \va{R}^{T}\va{s}
\end{equation}
Doing so has the added benefited of inverting a square matrix. Since $\vb{R}\vb{R^T}$ is still ill-conditioned, we again use the Moore-Penrose pseudo inverse:
\begin{equation}
    \va{f} = (\vb{R}^{T}\vb{R})^{\dagger}\va{R}^{T}\va{s}
\end{equation}

Numerous methods exists to solve ill-posed problems such as that described by the Fredholm Integral Equation of the First Kind illustrated in X-ray spectral analysis (equation \ref{eqn:spec}; e.g. \citealt{hansen_numerical_1992}). We explore two methods: preconditioning (e.g. \citealt{estatico_class_2002}) and regularization (e.g. \citealt{neumaier_solving_1998}). We apply standard preconditioning using the normal equations; however, the regularization technique is more involved.


\subsubsection{Regularization}
Matrix regularization is a family of algorithms designed to overcome ill-conditioned matrices by imposing a strict condition such as smoothness on a least-squared solution. We apply Tikhonov regularization which augments the standard least-squares formalism by a Lagrangian mutliplier (e.g. \citealt{calvetti_tikhonov_2004}):
\begin{equation}\label{eqn:TR1}
    \underset{\va{f}\in \va{\mathcal{R}}}{\argmin} \Big\{\norm{\vb{R}\va{f}-\va{s}}_{L_2} - \lambda\norm{\vb{D}}_{L_2}\Big\}
\end{equation}
where $\lambda>0$ is the regularization parameter, $\vb{D}$ is the regularization matrix, and $\norm{\vdot{}}_{L_2}$ is the L2 (or Euclidean) norm (\citealt{horn_matrix_2012}). A standard choice for the regularization matrix is the identity matrix, $\mathcal{I}$. We optimize the value of $\lambda$ using the standard L-curve analysis (
\citealt{kindermann_simplified_2019}). The result of minimizing equation \ref{eqn:TR1} is the following expression for the intrinsic spectrum:
\begin{equation}\label{eqn:TR-sol}
    \va{f}_\mu = [\vb{R}\vb{R^T}+\mu\vb{\mathcal{I}}]^\dagger [\vb{R^T}\va{s}]
\end{equation}
We must create a set of mock X-ray spectra in order to test the feasibility of solving the matrix equation (equation \ref{eqn:spec_mat}) by inversion or a least-squares method.

\subsection{Creation of Data}\label{sec:mock}
 We use the \texttt{sherpa} (v4.13) tool \texttt{fake\_pha} in order to create mock spectra with an approximate signal-to-noise ratio of 20. We take two \texttt{rmf} and \texttt{arf} files from different regions on ACIS-I3 from the observation 7253 (we note that the choice in observation is arbitrary and used to demonstrate the spatial changes in the convolved response matrix; the choice of chip is also arbitrary). We tested several chips and several different observations covering a range of \textit{Chandra} observation cycles. We report no difference in our results.
We test two intrinsic emission types: a simple 1-dimensional powerlaw (\texttt{powerlaw}) and an absorbed thermal plasma emission model (\texttt{phabs*apec}). The powerlaw index parameter is set to $-0.5$, the column density, $n_H$ is set to $10^{20}$cm$^{-2}$. Although these values are arbitrary, we tested several values with no change in the results.
Two spectra are created for each emission type; they differ only in the \arf{} and \rmf{} used in their creation.  

\section{Results \& Discussion} \label{sec:res}
When applied to the powerlaw model, the full spectral unfolding using Tikhonov regularization paired with a normalized preconditioner recovers the intrinsic spectra correctly up to several percent ($<$3\% errors). However, when applied to a physically motivated thermal model (such as \texttt{MEKAL} or \texttt{APEC}), the algorithm only successfully recovers the underlying continuum spectrum. Unfortunately, the method fails to fully capture the prominent emission lines (such as \textit{Fe K-$\alpha$}). Recovering the precise shape and peak amplitude of these emission lines is crucial for subsequent calculations of temperature and metallicity. Therefore, this method is ill-suited to solve the inverse problem posed in equation \ref{eqn:spec}. 
The authors are currently exploring the use of recurrent neural networks to solve the equation.

\bibliography{ChandraDeconvolution}{}
\bibliographystyle{aasjournal}

\end{document}